\documentclass[a4paper,12pt]{article}
\usepackage[cp1251]{inputenc}
\usepackage[russian]{babel}

\usepackage{amsmath}

\def\bb{\begin{equation}}
\def\ee{\end{equation}}

\begin{document}

\vskip 0.4cm
\title{ ``Quantizations'' of isomonodromic Нamiltonian Garnier system with two degrees of freedom}
\author{ D. P. Novikov, B. I. Suleimanov{\footnote{e-mail: bisul@mail.ru}}} 
\date{}
\maketitle

\begin{abstract}
We consruct solutions of analogues of the nonstationary Schr\"{o}dinger equation corresponding
to the polynomial isomonodromic Hamiltonian Garnier system with two degrees of freedom.
This solutions are obtained from solutions of systems of linear ordinary differential equations
whose compatibility condition is the Garnier system. This solutions upto explicit transform
also satisfy the Belavin — Polyakov — Zamolodchikov equations with four time variables
and two space variables.
\end{abstract}

\newpage

\vskip 0.4cm
\begin{center}
\Large{ ``Квантования'' изомонодромной гамильтоновой системы Гарнье с двумя степенями свободы}

\large{ Д. П. Новиков, Б. И. Сулейманов} 
\end{center}

\begin{abstract}
Построены решения аналогов временных уравнений Шредингера, соответствующих изомонодромной полиномиальной гамильтоновой 
системе Гарнье с двумя степенями свободы. Они задаются решениями линейных обыкновенных дифференциальных уравнений, условием совместности которых является данная система Гарнье. Эти решения с помощью явных замен сводятся также к решениям  четырехвременных, проcтранственно двумерных уравнений Белавина --- Полякова --- Замолодчикова.

\end{abstract}

\bigskip
Ключевые слова: уравнения Шредингера, гамильтоновость, изомонодромные деформации, система Гарнье, уравнения
 Белавина --- Полякова --- Замолодчикова, уравнения Пенлеве

\section{Введение} В теории обыкновенных дифференциальных уравнений (ОДУ) типа Пенлеве в последние годы выделилась тема связей между  линейными уравнениями метода изомонодромных деформаций (ИДМ), которые совместны на этих нелинейных ОДУ, с линейными дифференциальными уравнениями в частных производных квантовой механики и квантовой теории поля \cite{Difufa}---\cite{Contd}. 
Хронологически первой  по этой теме является, по-видимому, статья \cite{Difufa}, где был установлен следующий факт: 

шесть канонических ОДУ Пенлеве $\lambda''_{tt}=f_j(t,\lambda,\lambda'_t)\quad(j=1,\dots,6)$ эквивалентны гамильтоновым системам 
\begin{equation}\label{Hams}
\lambda'_t=H'_\mu(t,\lambda,\mu),\qquad \mu'_t=-H'_\lambda(t,\lambda,\mu)
\end{equation}
с такими гамильтонианами $H=H_j(t,\lambda,\mu)$, что решения уравнений ИДМ из \cite{Garn}
$$W_{xx}=P(t,\lambda(t),\mu(t);x)W,\qquad W_{t}=A(t,\lambda(t),\mu(t);x)W, $$
условием совместности которых являются соответствующие системы (\ref{Hams}), c помощью явных замен вида
$\Psi=W\exp(S(t,x))$
переводятся в решения уравнений 
\begin{equation}\label{murzp}\varepsilon \frac{\partial\Psi }{\partial t}=
H(t,x,\varepsilon\frac{\partial}{\partial x})\Psi \qquad(\varepsilon=1).\end{equation}
Правые части уравнений (\ref{murzp}), уже не содержащие зависимости от $\lambda(t)$ и $\mu(t)$,  при конкретном выборе очередности действий операторов умножения на переменную $x$  и дифференцирования  по ней задаются гамильтонианами $H=H_j(t,\lambda,\mu)$  $(j=1,6)$ 
гамильтоновых систем (\ref{Hams}). 
А другой выбор такой очередности, позволяет \cite{faa} эти шесть линейных эволюционных уравнений символически записать и как уравнения 
\begin{equation}\label{murzm}\varepsilon\frac{\partial\Psi }{\partial t}=
H(t,x,-\varepsilon\frac{\partial}{\partial x})\Psi.\end{equation}
Из квантовомеханических уравнений Шредингера, зависящих от постоянной Планка $h=2\pi\hbar=-2\pi i\varepsilon$, эволюционные уравнения (\ref{murzm}) получаются в результате формальной замены $\varepsilon=1$. Далее в этой статье такого сорта аналоги уравнения Шредингера, которые не зависят от постоянной Планка, мы, следуя терминологии, введенной в работе \cite{BTMF}, будем называть ``квантованиями'' соответствующих  гамильтоновых систем. 

{\Large Замечание 1.} ``Квантования'' (\ref{murzp}) с $\varepsilon=1$ встречаются в задачах диффузии \cite{Ovs},\cite{Dovs}. При исследовании некоторых проблем теоретиковероятностного  характера, описанная связь между ``квантованием'' (\ref{murzp}) ОДУ Пенлеве и соответствующими уравнениями ИДМ использовалась в  \cite{Buv}, \cite{Bluv} 
 и в серии публикаций Руманова \cite{Rum1}---\cite{Rum4}. Но в ряде частных cлучаев эта связь оказывается полезной \cite{mgu11}, \cite{Ufmgbis}, \cite{Nag}, \cite{Gran} и для построения решений квантовомеханических временных уравнений Шредингера (\ref{murzm}), в которых  $\varepsilon=i\hbar$. 
 
Связь линейных уравнений ИДМ с уравнениями квантовой теории поля была выявлена в \cite{Novd}. В этой статье показано, что совместные уравнения ИДМ \begin{equation}\label{sch1}
\Phi'_x=\sum_{i=1}^m\frac{A_i}{x-t_i}\Phi,\qquad \Phi'_{t_i}=-\frac{A_i}{x-t_i}\Phi
\end{equation}
для решений систем Шлезингера в матрицах $A_i$ размера $2\times2$ (системы Шлезингера 
\begin{equation}\label{sch3}
\frac{\partial A_j}{\partial t_i}=
\frac{[A_i,A_j]}{t_i-t_j},\quad i\neq j\;,\quad
\frac{\partial A_i}{\partial t_i}=-\sum_{j\neq i}\frac{[A_i,A_j]}{t_i-t_j}\quad (i,j=\overline{1,m}),
\end{equation}
были открыты \cite{schle} именно в качестве условия совместности ОДУ (\ref{sch1}))  
заменой  
$$\Psi=\tau(t_1,\dots,t_m)\Phi\qquad((\ln\tau)'_{t_i}=
\sum\limits_{j\neq i}^m
{\mbox{tr}A_jA_i}/(t_i-t_j)),$$
переводятся в систему ($\Delta_i$$=-\det A_i$, $\Delta_{\infty}$--- постоянные, $A_{\infty}$ --- постоянная матрица)
$$\sum\frac{\Psi'_{t_i}}{x-t_i}=\Psi''_{xx}-\sum\frac{\Delta_i \Psi}{(x-t_i)^2},
\sum(x-t_i)\Psi'_{t_i}=\Big(\sum\Delta_i-A_{\infty}-\Delta_{\infty}\Big)\Psi,
\sum\Psi'_{t_i}=-\Psi'_x.$$
При условии диагональности  матрицы  $A_{\infty}$ эти уравнения 
совпадают \cite{Novd} с пространственно одномерными уравнениями Белавина --- Полякова --- Замолодчикова (БПЗ) \cite{bel},\cite{FatZam} минимальной
модели двумерной квантовой теории поля с конформной группой симметрий,   
для которых центральный заряд алгебры Вирасоро, 
равен 1. В приложении B cтатьи \cite{Novd} также указано, что та же $\tau$-функция и cовместные решения $\Phi$ линейных ОДУ (\ref{sch1}) задают $2\times2$ 
  матрицы 
\begin{equation}\label{NovDP}M(x_1,x_2)=\tau
\Phi^{-1}(x_1)\Phi(x_2) \qquad (x_1=x,\quad x_2=y),\end{equation}
удовлетворяющие четырем линейным дифференциальным уравнениям в частных производных c дифференцированиями уже по двум пространственным независимым переменным $x_1$, $x_2$. 
Простой заменой (см. второй раздел настоящей статьи) эти четыре скалярных уравнения сводятся к пространственно двумерным уравнениям БПЗ.  

Формула  (\ref{NovDP}) (по ее поводу см. также  замену (2.3.36) из \cite{sato}) позднее послужила основой построения в \cite{faa} cовместных решений ''квантований'' 
\begin{equation}\label{quantd} \varepsilon \frac{\partial \Psi}{\partial t_i}=
H_{t_i}(t_1,t_2,x_1,x_2,-\varepsilon\frac{\partial}{\partial x_1},-\varepsilon\frac{\partial}{\partial x_2})\Psi \qquad( i=1,2),\end{equation} 
определяемых гамильтонианами $H_{t_i}(t_1,t_2,q_1, q_2, p_1, p_2)$ гамильтоновых систем 
$$(q_j)'_{t_i}=(H_{t_i})'_{p_j}, \qquad (p_j)'_{t_i}=-(H_{t_i})'_{q_j}\qquad(j=1,2),$$
которые представляют собой  изомонодромные высшие  аналоги первого и второго ОДУ Пенлеве: эти решения эволюционных уравнений (\ref{quantd}) через совместные решения соответствующих линейных уравнений ИДМ выражаются формулой типа (\ref{NovDP}). 

Естественным выглядит вопрос: можно ли таким образом образом строить решения (\ref{quantd}) для других изомонодромных гамильтоновых  систем с двумя степенями свободы? В настоящей статье справедливость положительного ответа на него демонстрируется для полиномиальной гамильтоновой системы Гарнье, которая была введена в рассмотрение в \cite{Kioko}, и которая c помощью явных преобразований \cite{Kioko} ---\cite{Kim}  сводится к давно известной изомонодромной системе Гарнье \cite{Garn}. 

{\Large Замечание 2.} Список \cite{Hsak}---\cite{Kanasa} известных на сегодня   гамильтоновых систем  с двумя степенями свободы, к которым применим ИДМ, с уверенностью полным считать нельзя. 
Но хорошо известно, что  система Гарнье возглавляет целую иерархию таких систем --- они  из системы Гарнье получаются \cite{Kim},\cite{Kanasa} процедурой последовательного вырождения. 
C помощью данной процедуры  приводимые ниже конструкции, вероятно, могут быть расширены и на всю эту иерархию 
(системы из \cite{faa} являются двумя низшими членами данной иерархии). 

В своих построениях мы будем cущественно опираться на упомянутые результаты \cite{Novd}. Их изложение и уточнение связи  этих результатов  с пространственно двумерными уравнениями БПЗ производится в  начале следующего раздела
 статьи.  Далее в разделе {\bf 2} показывается, что в случае четырех времен данные уравнения БПЗ
эквивалентны совместной системе эволюционных уравнений, два из которых есть ``квантования''  вида (\ref{quantd}) гамильтоновой изомонодромной системы Гарнье с двумя степенями свободы в форме, выписанной Окамото в \cite{Okomot} (ниже она называется системой Гарнье --- Окамото (ГО)). 
Эта система  связана  \cite{Okomot}, \cite{Gaup} с частным случаем системы  (\ref{sch3}) с четырьмя временами ($m=4$). Однако, как подчеркивается в заключительном пункте {\bf 2.5}  раздела {\bf 2} представляемой статьи, сведение всех решений  упомянутой системы ГО с двумя степенями свободы к решениям системам Шлезингера с четырьмя $t_m$  в явном виде до сих пор не было описано (о сложности и тонкости вопроса связи между системами Шлезингера и Гарнье, на которую указывалось еще cамим Гарнье \cite{Garn6},  можно составить представление по разделам 5.3 и 5.4 обзора \cite{babi}).  Между тем, во втором и третьем разделах нашей работы решения эволюционных уравнений (\ref{quantd}), определяемых общей системой ГО, строятся в терминах решений линейных ОДУ (\ref{sch1}), коэффициенты которых заданы именно всевозможными решениями общей системы Шлезингера (\ref{sch3}) с четырьмя  $t_m$.   

К последним же, как показано ниже в разделе {\bf3}, сводятся все решения 
полиномиальной системы Гарнье с двумя степенями свободы. В этом разделе выводится также формула (\ref{dno}), связывающая системы ГО и полиномиальные системы Гарнье для довольно широкого класса случаев.  Но, как продемонстрировано в конце раздела {\bf3},  однозначного соответствия между решениями этих двух систем данная простая формула во всех случаях не дает.

Зато с использованием ее квантового аналога --- замены  (\ref{qdno}) в заключительном разделе {\bf4}  статьи показывается, что пара уравнений (\ref{quantd}), определяемых общей системой ГО, эквивалентна совместной паре ``квантований'' (\ref{quantd}), задаваемой гамильтонианами общей полиномиальной системы Гарнье. И, таким образом, решения этих ``квантований'' выписываются в терминах совместных решений систем линейных ОДУ, коэффициенты  которых задаются как раз всевозможными решениями соответствующей полиномиальной системы Гарнье.

\section{ Cистемы Шлезингера, пространственно двумерные уравнения Белавина---Полякова---Замолодчи\-кова и ``квантования'' системы Гарнье --- Окамото}

{\bf2.1. }Вторую группу в системе 
уравнений Шлезингера (\ref{sch3}) заменяет условие
постоянства по $t_1,\ldots,t_m$  матрицы $A_1+\ldots+A_m=A_\infty$. Без фактического ограничения общности $2\times2$ системы Шлезингера рассматривались в \cite{Novd} в предположении, что их решения $B_i$
таковы, что: 
$\mbox{tr}B_i=0,$ $\det B_i=-\Delta_i$ $=-\theta_i^2/4$
 (постоянные $\pm \theta_i/2$ есть собственные числа $B_i$), $\det B_\infty=-\Delta_\infty,$ 
a матрица $B_\infty$ имеет один из двух видов:
\begin{equation}\label{Jordt}B_\infty=\frac12\begin{pmatrix}k_\infty&0\\0&-k_\infty\end{pmatrix},\end{equation}
\begin{equation}\label{Jordo}
B_\infty=\begin{pmatrix}0&0\\1&0\end{pmatrix}.
\end{equation}

{\bf 2.2.} В  пункте {\bf В1} приложения {\bf V}\cite{Novd} выписана система четырех 
уравнений 
$$\sum\left(\frac1{x-t_i}+\frac1{y-x}\right)M'_{t_i}
=M''_{xx}-\sum\frac{\Delta_i}{(x-t_i)^2}M,
$$
$$\sum\left(\frac1{y-t_i}+\frac1{x-y}\right)
M'_{t_i}=M''_{yy}-\sum\frac{\Delta_i}{(y-t_i)^2}M,$$
$$\sum M'_{t_i}+M'_x+M'_y=0,\quad
t_iM'_{t_i}+xM'_x+yM'_y=(\Delta_\infty-\sum\Delta_i)M,$$
решением которых является матрица (\ref{NovDP}).
Результат же $Y(t,x,y)$ замены
$$M=(x-y) \prod_{i=1}^{m} [(x-t_i)(y-t_i)]^{\theta_i/2}\exp{S(t)} Y \quad(S_{t_i}=\frac
{\theta_i}{2}\sum_{j\neq i}\frac
{\theta_j}{(t_i-t_j)}), $$
удовлетворяет пространственно двумерной сиcтеме БПЗ (см. систему 
(10) из \cite{Stoyan})
\begin{equation} \label{BPZ}
\sum_{i=1}^m\frac{Y'_{t_i}}{x-t_i}=Y''_{xx}+
\frac{Y_x-Y_y}{x-y}+\sum_{i=1}^m\frac{\theta_i}{x-t_i}Y'_x,\quad
\sum_{i=1}^m\frac{Y'_{t_i}}{y-t_i}=Y''_{yy}+
\frac{Y'_x-Y'_y}{x-y}+\sum_{i=1}^m\frac{\theta_i}{y-t_i}Y'_y,
\end{equation}
\begin{equation}\label{odn}
\sum_{i=1}^mY'_{t_i}+Y'_x+Y'_y=0,\qquad
\sum_{i=1}^mt_iY'_{t_i}+xY'_x+yY'_y=\lambda Y=(\Delta_{\infty}-(1+\sum_{i=1}^m\frac{\theta_i}{2})^2)Y.
\end{equation}

{\bf2.3.} Решения $B_i$ системы Шлезингера (\ref{sch3}) из пункта {\bf 2.1} сдвигами 
\begin{equation}\label{diag} Q_i=\begin{pmatrix}q_{11}^i(t)&q_{12}^i(t)\\q_{21}^i(t)&q_{22}^i(t)\end{pmatrix}=B_i+\frac{\theta_i}{2} \begin{pmatrix}1&0\\0&1\end{pmatrix}\end{equation}
переводятся в такие решения $Q_i$  этой же системы, что 

(i) cобственные значения матриц $Q_i(t)$ есть нуль и 
постоянная  $\theta_i$. 

Преобразование $Z= \Phi \prod_{i=1}^{4} (x-t_i)^{\theta_i/2} $ 
при этом решения $\Phi$ систем (\ref{sch1}) c матрицами $A_i=B_i$ переводят в решения $Z$ этих же систем, но   уже с матрицами $A_i=Q_i$.

Далее рассматриваются эти решения $Q_i$ уравнений Шлезингера (\ref{sch3}) в случае четырех $t_i$ c последующей фиксацией переменных $t_3$ и $t_4$ 
\begin{equation} \label{yard}  \qquad t_3=1, \qquad t_4=0.\end{equation}

{\Large Замечание 3.}
Эта фиксация  общности рассмотрения по сути не ограничивает. Действительно, 
совместные решения $A_k$ $(k=1,\dots, 4)$ уравнений (\ref{sch3}) с  независимыми переменными $t_1$ и $t_2$ и фиксацией (\ref{yard}) можно рассматривать как начальные данные для уравнений (\ref{sch3}) с независимой переменной $t_3$ при $t_3=1$.  Решение начальной задачи для уравнений Шлезингера (\ref{sch3}) с независимой переменной $t_3$ и этими начальными данными можно, в свою очередь, рассматривать в качестве начальных данных при  $t_4=0$ для системы уравнений (\ref{sch3}) с независимой переменной $t_4$. Решение же этой начальной задачи будет совместным решением всех уравнений (\ref{sch3}). 

(ii) в предположении, что решения $B_i$ системы Шлезингера (\ref{sch3}) из предыдущего раздела определяют постоянную матрицу $B_{\infty}$ вида (\ref{Jordt}), справедливо соотношение
\begin{equation}\label{Okomn}-\sum_{i=1}^4Q_i(t)=\begin{pmatrix}\chi&0\\0&\chi+\theta_\infty-1\end{pmatrix},\end{equation}
где $\theta_\infty=k_\infty+1$ от $t_i$ не зависит и $\chi=-\frac12(\sum_{i=1}^4\theta_i+\theta_\infty-1).$

Пусть еще эти решения $Q_i(t)$ системы Шлезингера удовлетворяют условиям: 

(iii)  элемент $q_{12}(x,t)$ матрицы 
$$Q(x,t)=\begin{pmatrix}q_{11}(x,t)&q_{12}(x,t)\\q_{21}(x,t)&q_{22}(x,t)\end{pmatrix}=\sum_{i=1}^{4}\frac{Q_i}{x-t_i}\label{simpe12},$$
очевидно, равный величине 
$$q_{12}(x,t)=\frac{X(t)x^2+a_1(t)x+a_2(t)}{T(x,t)}=\frac{\sum_{i=1}^4t_iq_{12}^ix^2+a_1(t)x+a_2(t)}{\prod_{i=1}^4 (x-t_i)},$$ 
таков, что функция $X(t)=\sum_{i=1}^4t_iq_{12}^i$ не есть тождественный нуль;

(iv) все нули $x(t)=\lambda_j(t)$ этого элемента $q_{12}(x,t)$ простые. 

Из предположений (ii) --- (iv) cледует \cite{Okomot} (cм. также 
раздел 6.2 книги \cite{Gaup}), что: 

a) $q_{12}(x,t)$ имеет два простых нуля и задается формулой
\begin{equation}\label{q12}
q_{12}(x,t)=\sum_{i=1}^{4}\frac{q_{12}^i(t)}{x-t_i}=X(t) \frac{\Lambda(x,t)}{T(x,t)}=X(t)\frac{(x-\lambda_1(t))(x-\lambda_2(t)}{T(x,t)}; \end{equation}

b) компонента $z(x,t)=z_1(x,t)$ любого вектора-решения $Z=(z_1,z_2)$
 первой из систем ОДУ (\ref{sch1}) с матрицами $A_i(t)=Q_i(t)$ удовлетворяет линейному ОДУ 
\begin{equation}\label{Garx}z''_{xx}=(\sum_{i=1}^4\frac{\theta_i-1}{x-t_i}+\sum_{k=1}^2\frac{1}{x-\lambda_k})z'_x
-(\frac{\kappa}{x(x-1)}-\sum_{i=1}^2\frac{t_i(t_i-1)K_i}{x(x-1)(x-t_i)}+\sum_{k=1}^2\frac{\lambda_k(\lambda_k-1)\mu_k}{x(x-1)(x-\lambda_k)})z,\end{equation}
\begin{equation}\label{Garxc}
\kappa=\frac{1}{4}[(\sum_{i=1}^4\theta_i-1)^2-\theta_{\infty}^2],
\quad \mu_{k}=
\sum_{i=1}^4\frac{q_{11}^i}{\lambda_k-t_i}.
\end{equation}

Совместность систем уравнений (\ref{sch1}) означает, что наряду с уравнением (\ref{Garx}) компонента $z(x,t)$ удовлетворяет также 
эволюционным уравнениям первого порядка 
\begin{equation}\label{Garto}z'_{t_i}=C_i(x,t)z'_x+J_i(x,t)z=s_i(t)\frac {T(x,t)}{(x-t_i) \Lambda (x,t)}z'_x+J_i(x,t)z\qquad(i=1,2).\end{equation}

В свою очередь, замена
$$z(x,t)=\prod_{i=1}^{4} (x-t_i)^{(\theta_i-1)/2}(x-\lambda_1(t))^{1/2}(x-\lambda_2(t))^{1/2}v(x,t)$$
это решение $z(x,t)$ уравнений (\ref{Garx}) и (\ref{Garto}) переводит в решение cистемы 
$$v''_{xx}=[\frac{c_3}{x^2}+\frac{c_4}{(x-1)^2}+\frac {c_5}{x(x-1)}
+\sum_{i=1}^2(\frac{c_i}{(x-t_i)^2}+\frac{\alpha_i(t)}{x(x-1)(x-t_i)})+$$
$$\sum_{j=1}^2(\frac{3}{4(x-\lambda_j(t))^2}+\frac{\beta_j(t)}{x(x-1)(x-\lambda_j(t))})]v\qquad((c_j)'_{t_i}=0),$$$$v'_{t_i}=C_i(x,t)v'_x+(-(C_i(x,t))'_x/2 +\gamma_i(t))v.
$$
 А, значит \cite{Garn}, \cite{Okomot}, условие совместности ОДУ (\ref{Garx}) c уравнением первого порядка (\ref{Garto}) есть  совместные между собой гамильтоновы системы с двумя степенями свободы 
\begin{equation}\label{kokgarl} \frac{\partial \lambda_k}{\partial t_j}=\frac{\partial K_j}{\partial \mu_k},\quad
\frac{\partial \mu_k}{\partial t_j}=-\frac{\partial K_j}{\partial \lambda_k}\quad(j,k=1,2),\end{equation}
где  $\lambda_i$  есть описанные выше нули элемента $q_{12}(x,t)$,  импульсы $\mu_i$ задаются формулами (\ref{Garxc}), а гамильтонианы $K_i$  формулами ($\delta_{mi}$ --- символ Кронекера)
\begin{equation}\label{kokgarK}
K_i=K_i(t_1,t_2,\lambda_1,\lambda_2,\mu_1,\mu_2;\theta_1,\theta_2,\theta_3,\theta_4,\kappa)=$$
$$=M_i\sum_{k=1}^2M^{k,i}[\mu_k^2-(\sum_{m=1}^2\frac{\theta_m-\delta_{im}}{\lambda_k-t_m}+\frac{\theta_3}{\lambda_k-1}+\frac{\theta_4}{\lambda_k})\mu_k+\frac{\kappa}{\lambda_k(\lambda_k-1)}],\end{equation}
$$M_i=
-\frac{(\lambda_1-t_i)(\lambda_2-t_i)}{(t_i-t_{i+1})(t_i-1)t_i},
\quad M^{k,i}=
\frac{(\lambda_k-t_{i+1})(\lambda_k-1)\lambda_k}{\lambda_k-\lambda_{k+1}}.$$
Уравнения (\ref{kokgarl}) 
и представляют собой изомонодромную гамильтонову систему ГО с  двумя степенями свободы  
\cite{Gaup} , \cite{MMaz}. 

{\Large Замечание 4.} На самом деле в \cite{Okomot} и в \cite{Gaup} условие (iii) явно не cформулировано. Зато там накладываются дополнительные ограничения на соответствующие матрицы $Q_i$. В частности, в начале раздела 6.2 \cite{Gaup}  приведено  следующее ограничение: 

(v) собственные значения матриц $Q_i$ не есть целые числа.

{\bf 2.4.} При m=4 и фиксации (\ref{yard}) 
уравнения  
(\ref{BPZ}) эквивалентны системам 
\begin{equation}\label{kevol1}
t_1(t_1-1)(t_1-t_2)Y'_{t_1}=\frac{(x-t_1)(y-t_1)(x-t_2)(x-1)x}
{y-x}[Y''_{xx}+$$
$$Y'_x(\frac{\theta_1}{x-t_1}+\frac{\theta_2+1}{x-t_2}+
\frac{\theta_3+1}{x-1}+\frac{\theta_4+1}{x})-\frac{\lambda}{x(x-1)}Y]-$$
$$\frac{(x-t_1)(y-t_1)(y-t_2)(y-1)y}
{y-x}[Y''_{yy}
+Y'_y(\frac{\theta_1}{y-t_1}+\frac{\theta_2+1}{y-t_2}+
\frac{\theta_3+1}{y-1}+\frac{\theta_4+1}{y
})-\frac{\lambda}{y(y-1)}Y],
\end{equation}
\begin{equation}\label{kevol2}
t_2(t_2-1)(t_2-t_1)Y'_{t_2}=\frac{(x-t_2)(y-t_2)(x-t_1)(x-1)x}
{y-x}[Y''_{xx}+$$
$$Y'_x(\frac{\theta_1+1}{x-t_1}+\frac{\theta_2}{x-t_2}+
\frac{\theta_3+1}{x-1}+\frac{\theta_4+1}{x})-\frac{\lambda}{x(x-1)}Y]-$$
$$\frac{(x-t_2)(y-t_2)(y-t_1)(y-1)y}
{y-x}[Y''_{yy}
+Y'_y(\frac{\theta_1+1}{y-t_1}+\frac{\theta_2}{y-t_2}+
\frac{\theta_3+1}{y-1}+\frac{\theta_4+1}{y})-\frac{\lambda}{y(y-1)}Y],
\end{equation} дополненных 
уравнениями (\ref{odn}). Уравнения  же (\ref{kevol1}), (\ref{kevol2}) ecть  ``квантования''
гамильтоновой системы ГО 
двух переменных (\ref{kokgarl})
: ввиду  операторных соотношений
$$\frac{\partial}{\partial x}x-x\frac{\partial}{\partial x}=1,\qquad \frac{\partial}{\partial y}y-y\frac{\partial}{\partial y}=1 $$
при подходящем выборе порядков действия операторов дифференцирования по переменным $x, y$ и умножения на многочлены этих переменных   данные эволюционные уравнения символически можно представить в виде  ($\varepsilon=1$)
\begin{equation}\label {qugo}
\varepsilon \frac{\partial Y}{\partial t_i}=K_i(t_1,t_2,x,y,-\varepsilon\frac{\partial}{\partial x},-\varepsilon\frac{\partial}{\partial y};\theta_1,\theta_2,\theta_3,\theta_4,\kappa)Y \qquad(i=1,2),
\end{equation}
где $K_i(t_1,t_2,\lambda_1,\lambda_2,\mu_1,\mu_2;\theta_1,\theta_2,\theta_3,\theta_4,\kappa)$ есть гамильтонианы (\ref{kokgarK}) этой гамильтоновой системы.  

{\Large Замечание 5}. При подходящем выборе порядков действия этих операторов уравнения  (\ref{kevol1}), (\ref{kevol2}) можно также представить  как ``квантования'' вида (\ref{qugo}) $(\varepsilon=1)$
$$\varepsilon \frac{\partial Y}{\partial t_i}=K_i(t_1,t_2,x,y,-\varepsilon\frac{\partial}{\partial x},-\varepsilon\frac{\partial}{\partial y};\hat{ \theta_1},\hat{ \theta_2},\hat{ \theta_3},\hat{ \theta_4},\hat{ \kappa})Y \qquad(i=1,2)$$
гамильтоновых систем ГО (\ref{kokgarl}),
которые вместо постоянных $\theta_k$, (k=1,4)  и $\kappa$ зависят от любых других постоянных $\hat{\theta_k} (k=1,4)$  и  $\hat{\kappa}$.
В частности, для всех гамильтонианов (\ref{kokgarK})
уравнения  (\ref{kevol1}), (\ref{kevol2}) можно символически записать и в виде $(\varepsilon=1)$
$$\varepsilon Y_{t_i}=K_i(t_1,t_2,x,y,\varepsilon\frac{\partial}{\partial x},\varepsilon\frac{\partial}{\partial y};\theta_1,\theta_2,\theta_3,\theta_4,\kappa)Y \qquad(i=1,2).$$

{\bf 2.5.} Выше приведена конструкция, которая позволяет явно выписать  совместные  решения эволюционных уравнений (\ref{kevol1}),(\ref{kevol2}) в терминах совместных решений линейных систем ОДУ  (\ref{sch1}) для $m=4$. Коэффициенты же последних задаются множеством решений  нелинейных систем Шлезингера (\ref{sch3}).  То есть, имеется однозначное соответствие этих совместных решений эволюционных уравнений (\ref{kevol1}),(\ref{kevol2}) данному множеству. Поэтому возникает естественное желание выразить это множество решений системы Шлезингера через решения гамильтоновой  системы ГО  (\ref{kokgarl}) с теми же независимыми переменными $t_1, t_2$. Однако  вопрос о соотношении между  системами Шлезингера и Гарнье не так прост.  Не совсем точны, например, высказываемые иногда утверждения о том, что в статье Окамото \cite{Okomot}  описана эквивалентность между этими системами. Даже в случае двух степеней свободы системы ГО, это  в \cite{Okomot} сделано лишь при справедливости для решений $B_i$ системы Шлезингера равенства (\ref{Jordt}) c ненулевыми постоянными $k_{\infty}$  и ряда других предположений. 
В частности, в этой статье не показано,  что все возможные решения системы ГО с двумя степенями свободы сводятся к решениям систем Шлезингера размера $2\times 2$. До сих пор в явном виде это, вообще, нигде не было не сделано. В следующем разделе показывается как к решениям систем Шлезингера размера $2\times 2$ без всяких дополнительных предположений могут быть сведены общие решения полиномиальной гамильтоновой системы Гарнье с двумя степенями свободы. 
Во многом этот раздел основан на результатах недавних работ \cite{Hsak}, \cite{Kanasa}. Но в пунктах {\bf3.3}---{\bf3.5} излагаются  и довольно существенные дополнения к этим результатам.

\section{Полиномиальная система Гарнье} 

{\bf3.1} Полиномиальную систему Гарнье составляют совместные гамильтоновы системы 
\begin{equation}\label{Kanaso} (q_j)'_{t_i}=(H_{Gar,t_i})'_{p_j},\qquad  (p_j)'_{t_i}=-(H_{Gar,t_i})'_{q_j} \qquad(i,j=1,2)\end{equation}
с гамильтонианами $H_{Gar,t_i}$ и $H_{Gar,t_{i+1}}$, 
первый из которых  задается формулой 
\begin{equation}\label{Hkanasa} t_i(t_i-1)H_{Gar,t_i}=q_i(q_i-1)(q_i-t_i)p_i^2+[(\theta^0+\theta^{t_{i+1}}+1)q_i(q_i-1)-$$
$$(2\theta_2^{\infty}+\theta^1+\theta^0+\theta^{t_i}+\theta^{t_{i+1}}+1)q_i(q_i-t_i)+\theta^{t_i}(q_i-1)(q_i-t_i)]p_i+\theta_2^{\infty}(\theta_2^{\infty}+\theta^1)q_i +$$
$$(2q_ip_i+q_{i+1}p_{i+1}-\theta^1-2\theta_2^{\infty})q_iq_{i+1}p_{i+1}-
\frac{1}{t_i-t_{i+1}}[t_i(t_i-1)(p_iq_i+\theta^{t_i})p_iq_{i+1}-$$
$$t_i(t_{i+1}-1)(2p_iq_i+\theta^{t_i})p_{i+1}q_{i+1}+t_{i+1}(t_i-1)q_i(p^2_{i+1}q_{i+1}+
\theta^{t_{i+1}}(p_{i+1}-p_i))],\end{equation}
а второй получается из (\ref{Hkanasa}) заменой   $t_1\leftrightarrow t_2$,
$q_1\leftrightarrow q_2,$ $p_1\leftrightarrow p_2.$
В этих гамильтонианах 
$\theta^0,$ $\theta^{1},$ $\theta^{t_{1}},$ $\theta^{t_{2}},$ $\theta_1^{\infty},$ $\theta_2^{\infty}$
есть постоянные, которые связаны так называемым соотношением Фукса (соответствующим выписываемой ниже системе 
ИДМ (\ref{isoms}))
\begin{equation}\label{Fuxc} \theta^0+\theta^{1}+\theta^{t_{1}}+\theta^{t_{2}}+\theta_1^{\infty}+\theta_2^{\infty}=0.\end{equation}

Первая из гамильтоновых систем (\ref{Kanaso}) представляет собой  систему  ОДУ 
\begin{equation} \label{oqo}t_i(t_i-1)(q_i)'_{t_i}=2p_iq_i[(q_i-1)(q_i-t_i)-\frac{t_i(t_i-1)}{t_i-t_{i+1}}q_{i+1}]+
2p_{i+1}q_iq_{i+1}[q_i+\frac{t_i(t_{i+1}-1)}{t_i-t_{i+1}}]-$$ $$(\theta^1+2\theta_2^{\infty})q_i^2-(1+\theta^0+\theta^{t_i}+\theta^{t_{i+1}})q_i+(1+\theta^1+2\theta_2^{\infty}+\theta^0+\theta^{t_{i+1}})t_iq_i+
t_i\theta^{t_i}+$$
$$\frac{(t_i-1)}{t_i-t_{i+1}}[t_{i+1}\theta^{t_{i+1}}q_i-t_{i}\theta^{t_i}q_{i+1}],\end{equation}
\begin{equation} \label{opo}t_i(t_i-1)(q_{i+1})'_{t_i}=2p_{i}q_iq_{i+1}[q_i+\frac{t_i(t_{i+1}-1)}{t_i-t_{i+1}}]+
2p_{i+1}q_iq_{i+1}[q_{i+1}-\frac{t_{i+1}(t_{i}-1)}{t_i-t_{i+1}}]-$$ $$(\theta^1+2\theta_2^{\infty})q_iq_{i+1}-\frac{1}{t_i-t_{i+1}}[t_{i+1}(t_i-1)\theta^{t_{i+1}}q_i-t_{i}(t_{i+1}-1)\theta^{t_i}q_{i+1}],\end{equation}
\begin{equation} \label{oppo}t_i(t_i-1)(p_i)'_{t_i}=-p_i^2[3q_i^2-2(t_i+1)q_i+t_i-\frac{t_i(t_i-1)}{t_i-t_{i+1}}q_{i+1}]
-2p_{i+1}p_iq_{i+1}[2q_i+\frac{t_i(t_{i+1}-1)}{t_i-t_{i+1}}]-$$ $$p_{i+1}^2q_{i+1}[q_{i+1}-\frac{t_{i+1}(t_{i}-1)}{t_i-t_{i+1}}] +
p_i[ 2(\theta^1+2\theta_2^{\infty})q_i+(1+\theta^0+\theta^{t_i}+\theta^{t_{i+1}})-$$ $$(1+\theta^1+2\theta_2^{\infty}+\theta^0+\theta^{t_{i+1}})t_i
-\frac{t_{i+1}(t_i-1)\theta^{t_{i+1}}}{t_i-t_{i+1}}]+$$
$$p_{i+1}[(\theta^1+2\theta_2^{\infty})q_{i+1}+\frac{t_{i+1}(t_i-1)\theta^{t_{i+1}}}{t_i-t_{i+1}}]-\theta_2^{\infty}(\theta_2^{\infty}+\theta^1),\end{equation}
\begin{equation} \label{opt}t_i(t_i-1)(p_{i+1})'_{t_i}=p_i^2q_i\frac{t_i(t_i-1)}{t_i-t_{i+1}}
-2p_{i+1}p_iq_{i}[q_i+\frac{t_i (t_{i+1}-1)}{t_i-t_{i+1}}]-$$ $$p_{i+1}^2q_i[2q_{i+1}-\frac{t_{i+1}(t_{i}-1)}{t_i-t_{i+1}}] +p_i\frac{\theta^{t_i}t_i(t_i-1)}
{t_i-t_{i+1}}+p_{i+1}[(\theta^1+2\theta_2^{\infty})q_{i}-\frac{t_{i}(t_{i+1}-1)\theta^{t_i}}{t_i-t_{i+1}}],\end{equation}
а вторая систему  ОДУ
\begin{equation} \label{tqo}t_{i+1}(t_{i+1}-1)(q_{i})'_{t_{i+1}}=2p_{i}q_iq_{i+1}[q_i+\frac{t_i(t_{i+1}-1)}{t_i-t_{i+1}}]
+2p_{i+1}q_iq_{i+1}[q_{i+1}-\frac{t_{i+1}(t_{i}-1)}{t_i-t_{i+1}}]-$$ $$(\theta^1+2\theta_2^{\infty})q_iq_{i+1}-\frac{1}{t_i-t_{i+1}}[t_{i+1}(t_i-1)\theta^{t_{i+1}}q_i-t_{i}(t_{i+1}-1)\theta^{t_i}q_{i+1}],\end{equation}
\begin{equation} \label{tqt}t_{i+1}(t_{i+1}-1)(q_{i+1})'_{t_{i+1}}=2p_iq_iq_{i+1}[q_{i+1}-\frac{t_{i+1}(t_{i}-1)}{t_i-t_{i+1}}]+$$
$$2p_{i+1}q_{i+1}[(q_{i+1}-1)(q_{i+1}-t_{i+1})+\frac{t_{i+1}(t_{i+1}-1)}{t_i-t_{i+1}}q_{i}]-(\theta^1+2\theta_2^{\infty})q_{i+1}^2-$$
$$(1+\theta^0+\theta^{t_i}+\theta^{t_{i+1}})q_{i+1}+(1+\theta^1+2\theta_2^{\infty}+\theta^0+\theta^{t_i})t_{i+1}q_{i+1}+
t_{i+1}\theta^{t_{i+1}}+$$
$$\frac{(t_{i+1}-1)}{t_i-t_{i+1}}[t_{i+1}\theta^{t_{i+1}}q_i-t_{i}\theta^{t_i}q_{i+1}],\end{equation}
\begin{equation} \label{tpo}t_{i+1}(t_{i+1}-1)(p_{i})'_{t_{i+1}}=-p_{i}^2q_{i+1}[2q_{i}+\frac{t_{i}(t_{i+1}-1)}{t_i-t_{i+1}}] -2p_{i+1}p_iq_{i+1}[q_{i+1}-\frac{t_{i+1}(t_{i}-1)}{t_i-t_{i+1}}]-$$ $$p_{i+1}^2q_{i+1}\frac{t_{i+1}(t_{i+1}-1)}{t_i-t_{i+1}} +
p_{i}[(\theta^1+2\theta_2^{\infty})q_{i+1}+\frac{t_{i+1}(t_{i}-1)\theta^{t_{i+1}}}{t_i-t_{i+1}}]-p_{i+1}\frac{\theta^{t_{i+1}}t_{i+1}(t_{i+1}-1)}
{t_i-t_{i+1}},\end{equation}
\begin{equation} \label{tpt} t_{i+1}(t_{i+1}-1)(p_{i+1})'_{t_{i+1}}=-p_{i}^2q_{i}[q_{i}+\frac{t_{i}(t_{i+1}-1)}{t_i-t_{i+1}}]-2p_{i+1}p_iq_{i}[2q_{i+1}-\frac{t_{i+1}(t_{i}-1)}{t_i-t_{i+1}}]-
$$$$p_{i+1}^2[3q_{i+1}^2-2q_{i+1}(t_{i+1}+1)+t_{i+1}+\frac{t_{i+1}(t_{i+1}-1)}{t_i-t_{i+1}}q_{i}]+p_{i}[(\theta^1+2\theta_2^{\infty})q_{i}-\frac{t_{i}(t_{i+1}-1)\theta^{t_{i}}}{t_i-t_{i+1}}]+$$
 $$p_{i+1}[ 2(\theta^1+2\theta_2^{\infty})q_{i+1}+(1+\theta^0+\theta^{t_i}+\theta^{t_{i+1}})-(1+\theta^1+2\theta_2^{\infty}+\theta^0+\theta^{t_i})t_{i+1}
+\frac{t_{i}(t_{i+1}-1)\theta^{t_{i}}}{t_i-t_{i+1}}]-$$
$$\theta_2^{\infty}(\theta_2^{\infty}+\theta^1).\end{equation}

{\bf 3.2} При дополнительном предположении
\begin{equation}\label{neqg}\theta_1^{\infty}\neq \theta_2^{\infty}\end{equation}
одновременная справедливость гамильтоновых систем (\ref{Kanaso}), 
согласно \cite{Kanasa} есть  условие совместности  систем линейных ОДУ Шлезингера
\begin{equation}\label{isoms}
\frac{\partial Z}{\partial x}=\left(
\frac{S_0}{x}+\frac{S_1}{x-1}+\frac{S_{t_1}}{x-t_1}+\frac{S_{t_2}}{x-t_2}\right)Z,
\quad\frac{\partial Z}{\partial t_1}=-\frac{S_{t_1}}{x-t_1}Z, \quad
\frac{\partial Z}{\partial t_2}=-\frac{S_{t_2}}{x-t_2}Z.
\end{equation}
Здесь $S_0$, $S_1$, и $S_{t_i}$ есть матрицы, которые задаются формулами
\begin{equation} \label{align*}
S_{\xi}=
\begin{pmatrix}
1 &  0\\
0 & u
\end{pmatrix}^{-1}
P^{-1}\hat{A}_{\xi}P
\begin{pmatrix}
1 & 0\\
0 & u
\end{pmatrix} \qquad 
(\xi=0,1,t_1,t_2),\\
\end{equation}
\begin{equation} \label{align**}
\hat{A}_0=
\begin{pmatrix}
\theta^0 &-1+\dfrac{q_1}{t_1}+\dfrac{q_2}{t_2}\\
0 &0
\end{pmatrix},\quad
\hat{A}_1=
\begin{pmatrix}
\theta^1+\theta^\infty_2-p_1q_1-p_2q_2& 1 \\
\begin{matrix}
(p_1q_1+p_2q_2-\theta^\infty_2)\times\\
\times(\theta^1+\theta^\infty_2-p_1q_1-p_2q_2)
\end{matrix}
&p_1q_1+p_2q_2-\theta^\infty_2
\end{pmatrix},$$
$$\hat{A}_{t_i}=
\begin{pmatrix}
\theta^{t_i}+p_iq_i& -\dfrac{q_i}{t_i} \\
t_ip_i(\theta^{t_i}+p_iq_i)& -p_iq_i
\end{pmatrix},
\end{equation}
\begin{equation} \label{align***}
\quad P=
 \begin{pmatrix}
 1 & 0 \\
 \frac{a}{\theta^{\infty}_1-\theta^{\infty}_2} & 1
 \end{pmatrix} ,
\end{equation}
функция 
\begin{equation}\label{twoo}a=(p_1q_1+p_2q_2-\theta_2^{\infty})
(p_1q_1+p_2q_2-\theta^1-2\theta_2^{\infty})-t_1p_1(\theta^{t_1}+p_1q_1)-t_2p_2(\theta^{t_2}+p_2q_2)
\end{equation} есть элемент ``21'' матрицы $$\hat{A}_\infty=-\hat{A}_0-\hat{A}_1-\hat{A}_{t_1}-\hat{A}_{t_2}=\begin{pmatrix}
-\theta^0-\theta^{1}-\theta^{t_{1}}-\theta^{t_{2}}-\theta_2^{\infty}&0\\
a &\theta^\infty_2\end{pmatrix},$$
 а функция $u=u(t_1,t_2)$ --- cовместное решение двух дифференциальных уравнений
\begin{equation}\label{alogn}
t_i(t_i-1)\frac{1}{u}\frac{\partial u}{\partial t_i}=
q_i\{2p_i(t_i-q_i)+\theta^1+2\theta^\infty_2\}-2q_ip_{i+1}q_{i+1}+t_i\theta^{t_i} \quad(i=1,2).
\end{equation}

Хорошо видно, что собственные числа у каждой из матриц $\hat{A_{\xi}}$ 
равны 0 и $\theta^{\xi}$.  А их сумма с учетом соотношение Фукса (\ref{Fuxc}) 
задается формулой
$$\hat{A}_\infty:=-\hat{A}_0-\hat{A}_1-\hat{A}_{t_1}-\hat{A}_{t_2}
=\begin{pmatrix}
\theta^\infty_1&0\\
a &\theta^\infty_2\end{pmatrix}.$$
После преобразования (\ref{align*}) 
мы получаем решения 
\begin{equation} \label{Schs}Q_1=S_{t_1},\qquad Q_2=S_{t_2}, \qquad Q_3=S_1, \qquad Q_4=S_0\end{equation}
уравнений Шлезингера  (\ref{sch3}) при $m=4$ с фиксацией (\ref{yard})  и постоянными 
\begin{equation}\label{conss}\theta_1=\theta^{t_1},\qquad\theta_2=\theta^{t_2},\qquad \theta_3=\theta^1, \qquad \theta_4=\theta^0. \end{equation}
Их сумма задает постоянную матрицу
$$S_\infty=-S_0-S_1-S_{t_1}-S_{t_2}
=\begin{pmatrix}
\theta^\infty_1&0\\
0 &\theta^\infty_2\end{pmatrix}.$$
Ее равенство матрице (\ref{Okomn}) означает,  что в \cite{Kanasa} разбираются лишь решения cистемы Шлезингера, которые заменами (\ref{diag}) переводятся в множество решений этих систем, отвечающих  случаю (\ref{Jordt}) матрицы $B_{\infty}$ c отличной от нуля постоянной $k_{\infty}=\theta_2^{\infty}-\theta_1^{\infty}.$

{\bf3.3} При  
\begin{equation}\label{nojap}\theta_1^{\infty}=\theta_2^{\infty}\end{equation}
преобразование
$$S_{\xi}=\begin{pmatrix}
1 &  0\\
g_{21} & u
\end{pmatrix}^{-1}\hat A_{\xi} \begin{pmatrix}
1 &  0\\
g_{21} & u\end{pmatrix}
$$
с функцией $g_{21}$, удовлетворяющей совместной паре соотношений  
$$(g_{21})'_{t_k}-g_{21}u'_{t_k}=-p_i(\theta^{t_k}+p_kq_k) \qquad(k=i,i+1),$$
матрицы (\ref{align**}) переводят в 
решения  (\ref{Schs}) уравнений Шлезингера (\ref{sch3}), cумма которых имеет вид жордановой клетки:
$$-\sum_{i=1}^4Q_i(t)=\begin{pmatrix}\theta_1^{\infty}\\\nu&\theta_1^{\infty}\end{pmatrix}.
$$
Здесь $\nu$ есть постоянная, которая формулой 
$a(t_1,t_2)=\nu u(t_1,t_2)$ 
задает cовместное решение уравнений (\ref{alogn}) через функцию (\ref{twoo}). 
При этом собственные значения матриц $S_i$ --- по прежнему $0$ и постоянные $\theta_i$. После
переобозначения (\ref{conss}) мы получим решения (\ref{Schs}) cистем Шлезингера  (\ref{sch3}) при $m=4$ и фиксации (\ref{yard}), сумма которых с учетом сдвигов (\ref{diag}) задается в точности формулой 
$$B_\infty=\begin{pmatrix}0&0\\\nu&0\end{pmatrix}.$$

В случае $a\neq0$  можно положить $\nu=1$ и получить случай этих решений с нормировкой (\ref{Jordo}). Если же $a=0$ (мы не останавливаемся на вопросе  о совместности этого равенства с полиномиальной  системой Гарнье, поскольку он для основных целей нашей статьи не принципиален), то, положив $\nu=0$, получим соответствущие решения системы Шлезингера, редуцирующейся к случаю шестого ОДУ Пенлеве: см. Приложение  к этой статье.

{\bf 3.4} 
Полиномиальная гамильтонова система Гарнье (\ref{Kanaso})
 была первоначально выписана  в \cite{Kioko}. Там она с помощью явного преобразования
 выведена из системы ГО (\ref{kokgarl}),(\ref{kokgarK}).  При этом как постоянные, так и переменные  гамильтоновых систем ГО из работ \cite{Kioko} и \cite{Hsak},\cite{Kanasa} отличаются.  
 H.Sakai в \cite{Hsak} указывает, что решения этих систем ГО связаны преобразованием  Бэклунда, не приводя, однако, его явно. 
 
Мы же со своей стороны отмечаем здесь,  что при условии  справедливости неравенства (\ref{neqg}) и дополнительных ограничений (iii),(iv) решения полиномиальной  системы Гарнье (\ref{Kanaso}) 
 и решения системы ГО (\ref{kokgarl}),(\ref{kokgarK}) связаны простыми формулами:
\begin{equation}\label{dn1}\lambda_1+\lambda_2=\frac{t_1+t_2-(1+t_2)q_1-(1+t_1)q_2}{1-q_1-q_2},
\qquad \lambda_1\lambda_2=\frac{t_1t_2-t_2q_1-t_1q_2}{1-q_1-q_2},\end{equation}
\begin{equation}\label{dno}q_1=\frac{(1-t_2)(\lambda_1-t_1)(\lambda_2-t_1)}
{(t_1-t_2)(\lambda_1-1)(\lambda_2-1)},
\qquad q_2=-\frac{(1-t_1)(\lambda_1-t_2)(\lambda_2-t_2)}
{(t_1-t_2)(\lambda_1-1)(\lambda_2-1)}.\end{equation}
Действительно, посчитав элемент ``12'' в матрице коэффициентов
\begin{equation}\label{otwo} \frac{\hat A_0}{x}+\frac{\hat A_1}{x-1}+\frac{\hat A_{t_1}}{x-t_1}
+\frac{\hat A_{t_2}}{x-t_2},\end{equation}
 получим, что он равен
\begin{equation} 
\label{gaup}\frac{(1-q_1-q_2)x^2+[-t_1-t_2+q_2(1+t_1)+q_1(1+t_2)]x+t_1t_2-q_1t_2-q_2t_1}
{x(x-1)(x-t_1)(x-t_2)}.\end{equation} 
Поэтому равенства (\ref{dn1}),(\ref{dno}) следуют из формулы (\ref{q12}) и из того факта, что преобразование подобия (\ref{align*}) с матрицей (\ref{align***}) элемент ``12'' матриц не меняет.

Дифференцированием в силу систем (\ref{Kanaso}) и 
(\ref{kokgarl}) -- (\ref{kokgarK}) можно проверить, что при этом справедливы соотношения
$$p_1+\frac{\theta^{t_1}}{q_1}=\frac{(\lambda_1-1)(\lambda_2-1)}{(t_1-1)(t_2-1)}
\Big(\frac{(\lambda_1-1)(\lambda_1-t_2)\mu_1-(\lambda_2-1)(\lambda_2-t_2)\mu_2}{\lambda_1-\lambda_2}
-$$
$$(\theta^{t_1}+\theta^{t_2}+\theta^1+\theta^0+\theta^\infty_2)+
\frac{\theta^0t_2}{\lambda_1\lambda_2}\Big),$$
$$p_2+\frac{\theta^{t_2}}{q_2}=\frac{(\lambda_1-1)(\lambda_2-1)}{(t_1-1)(t_2-1)}
\Big(\frac{(\lambda_1-1)(\lambda_1-t_1)\mu_1-(\lambda_2-1)(\lambda_2-t_1)\mu_2}{\lambda_1-\lambda_2}
-$$
$$(\theta^{t_1}+\theta^{t_2}+\theta^1+\theta^0+\theta^\infty_2)+
\frac{\theta^0t_1}{\lambda_1\lambda_2}\Big).$$

{\bf 3.5} Однако это наше замечание вопрос об идентификации системы ГО (\ref{kokgarl}), (\ref{kokgarK}) с полиномиальной системой Гарнье (\ref{Kanaso}) из работ \cite{Hsak}, \cite{Kanasa}, описанной в этом разделе, полностью, конечно, не решает. 

Возьмем, например, такую редукцию полиномиальной сиcтемы Гарнье (\ref{Kanaso})
\begin{equation}\label{redH} q_i+q_{i+1}=1,\end{equation}
для которой формулы (\ref{dn1}) теряют смысл. В случае  
\begin{equation}\label{kond}\theta_1^{\infty}=\theta_2^{\infty}+1\end{equation}
такая редукция существует. В самом деле, cложение уравнений  
(\ref{oqo}) и (\ref{opo}) дает соотношение 
\begin{equation} \label{sumq}t_i(t_i-1)(q_i+q_{i+1})'_{t_i}=(q_i+q_{i+1}-1)[2p_iq_i(q_i-t_i)+2p_{i+1}q_{i+1}q_i-(\theta^1+2\theta_2^{\infty})q_i-t_i\theta^{t_
i}]+$$
$$(1+\theta^1+2\theta_2^{\infty}+\theta^0+\theta^{t_i}+\theta^{t_{i+1}})(t_i-1)q_i,\end{equation}
 которое в предположении справедливости  (\ref{redH}) влечет за собой тождество 
\begin{equation} \label{redt}(1+\theta^1+2\theta_2^{\infty}+\theta^0+\theta^{t_i}+\theta^{t_{i+1}})=0\end{equation}
(редукции $q_i=0$, $q_{i+1}=1$ cистема (\ref{Kanaso}) не выдерживает ни при каких значениях  $\theta^{\xi}$). 
Равенства же (\ref{kond}) и (\ref{redt}) с учетом соотношения Фукса (\ref{Fuxc}) равносильны. 

Правые части уравнений (\ref{opo}) и (\ref{tqo} ) совпадают. Поэтому из равенства (\ref{redH}) cледует также тождество
$t_i(t_i-1)(q_i)'_{t_i}+t_{i+1}(t_{i+1}-1)(q_i)'_{t_{i+1}}=0,$
означающее, что функция $q_i$ от своих аргументов зависит следующим образом 
\begin{equation}\label{autos}q_i(t_i,t_{i+1})=q_i(\omega), \qquad \omega=\frac{t_{i}(t_{i+1}-1)}{t_{i+1}-t_{i}}.\end{equation} 

При выполнении равенств  (\ref{redH}) и (\ref{kond}) результаты сложения уравнения (\ref{oppo}) с уравнением  (\ref{tpo}) 
и уравнения (\ref{opt}) c уравнением  (\ref{tpt}) cовпадают между собой: 
$$t_i(t_i-1)(p_i)'_{t_i}+t_{i+1}(t_{i+1}-1)(p_i)'_{t_{i+1}}=
t_i(t_i-1)(p_{i+1})'_{t_i}+t_{i+1}(t_{i+1}-1)(p_{i+1})'_{t_{i+1}}.$$
И, стало быть,  разность $p_i-p_{i+1}$ также зависит лишь от независмой переменной $\omega$: 
\begin{equation}\label{difg} P(t_i,t_{i+1})=p_i(t_i,t_{i+1})-p_{i+1}(t_i,t_{i+1})=P(\omega).\end {equation}
В силу (\ref{autos}) и (\ref{difg}) уравнение (\ref{opo}) и разность уравнения (\ref{oppo}) c уравнением (\ref{opt}) принимают при этом вид 
гамильтоновой системы 
\begin{equation} \label{fpg6} Q'_{\omega}=H'_P(\omega,Q,P),\qquad P'_{\omega}=-H'_P(\omega,Q,P), \end{equation}
где $Q=q_i(\omega)$, а гамильтониан $H$ имеет вид 
$$H=\frac{1}{\omega(\omega-1)}\{P^2Q(Q-1)(Q-\omega)-P[(\theta_1+2\theta_2^{\infty})Q(Q-1)+\omega\theta^{t_i}(Q-1)+(\omega-1)\theta^{t_{i+1}}Q]+$$
$$\theta_2^{\infty}(\theta_2^{\infty}+\theta^1)Q\}.$$
За исключением вырожденных решений этой гамильтоновой системы с $Q(\omega)=const$ и $Q(\omega)=\omega$ (cуществующих при дополнительных ограничениях на постоянные $\theta^{\xi}$), после исключения из 
(\ref{fpg6}) импульса $P(\omega)$ координата $Q(\omega)$ удовлетворяет общему случаю шестого ОДУ Пенлеве --- гамильтониан $H$ совпадает с известным \cite{Okok} полиномиальным гамильтонианом для этого ОДУ. После нахождения
$(Q(\omega),P(\omega))$ определение $p_i$ и $p_{i+1}$ cводится к решению совместных между собой уравнений Риккати.

При редукции (\ref{redH})  из вида (\ref{gaup}) элемента  ``12'' матрицы (\ref{otwo}) видно, что предположение (iii) предыдущего раздела о решениях соответствующих систем Шлезингера не выполнено (равенство (\ref{kond}), заметим, противоречит также ограничению (v) из \cite{Gaup}, процитированному в конце раздела {\bf3}). И, значит, вышеописанные конструкции  из \cite{Okomot}, \cite{Gaup}   сведения всех решений систем ГО к системам Шлезингера не описывают, даже если матрица $B_{\infty}$ для последних имеет вид (\ref{Jordt}) с $k_{\infty}\neq 0$. 

\section{``Квантования'' полиномиальной системы Гарнье}

Квантовый аналог формулы связи (\ref{dno})
 --- замена 
 \begin{equation}\label{qdno}\zeta=\frac{(1-t_2)(x-t_1)(y-t_1)}
{(t_1-t_2)(x-1)(y-1)},
\qquad \eta=-\frac{(1-t_1)(x-t_2)(y-t_2)}
{(t_1-t_2)(x-1)(y-1)}\end{equation}
и  явная замена 
\begin{equation}\label{chenqz}
Y=(xy)^{\alpha}(x-1)^{\beta}(y-1)^\beta V,
\end{equation}
где $\alpha$ и $\beta$ удовлетворяют равенствам 
$$\alpha(\alpha+\theta_4)=0,\quad
\lambda=2\alpha\beta+(\theta_3+1)\alpha+(\theta_4+1)\beta+\beta(\beta+\theta_3)+(\theta_1+\theta_2+1)(\alpha+\beta),$$
решения ``квантований'' (\ref{kevol1}), (\ref{kevol2}) 
переводят в совместные решения уравнений
\begin{equation}\label{quoopp}t_1(t_1-1)V'_{t_1}=[\zeta^3-(t_1+1)\zeta^2+t_1\zeta-\frac{t_1(t_1-1)\zeta\eta}{t_1-t_2}]V''_{\zeta\zeta}+$$
$$[2\zeta^2\eta+\frac{2t_1(t_2-1)\zeta\eta}{t_1-t_2}]V''_{\zeta\eta}+[\zeta\eta^2-\frac{t_2(t_1-1)\zeta\eta}{t_1-t_2}]V''_{\eta\eta}+$$
$$[-(\theta_3+2\beta-1)\zeta^2+t_1\zeta(\theta_2+\theta_3+\theta_4+2\alpha+2\beta)-\zeta(\theta_1+\theta_2+\theta_4+2\alpha+2)+$$
$$t_1(\theta_1+1)-\frac{(\theta_1+1)t_1(t_1-1)\eta}{t_1-t_2}+\frac{(\theta_2+1)t_2(t_1-1)\zeta}{t_1-t_2}]V'_{\zeta}+$$
$$[-(\theta_3+2\beta-1)\zeta\eta+\frac{(\theta_1+1)t_1(t_2-1)\eta}{t_1-t_2}-\frac{(\theta_2+1)t_2(t_1-1)\zeta}{t_1-t_2}]V'_{\eta}+$$
$$[\beta(\beta+\theta_3)\zeta+(t_1-1)\theta_1\alpha+t_1\theta_1\beta]V,
\end{equation}
\begin{equation}\label{quotpp}t_2(t_2-1)V'_{t_2}=[\eta^3-(t_2+1)\eta^2+t_2\eta+\frac{t_2(t_2-1)\zeta\eta}{t_1-t_2}]V''_{\eta\eta}+$$
$$[2\eta^2\zeta-\frac{2t_2(t_1-1)\zeta\eta}{t_1-t_2}]V''_{\zeta\eta}+[\eta\zeta^2+\frac{t_1(t_2-1)\zeta\eta}{t_1-t_2}]V''_{\zeta\zeta}+$$
$$[-(\theta_3+2\beta-1)\eta^2+t_2\eta(\theta_1+\theta_3+\theta_4+2\alpha+2\beta)-\eta(\theta_1+\theta_2+\theta_4+2\alpha+2)+$$
$$t_2(\theta_2+1)+\frac{(\theta_2+1)t_2(t_2-1)\zeta}{t_1-t_2}-\frac{(\theta_1+1)t_1(t_2-1)\eta}{t_1-t_2}]V'_{\eta}+$$
$$[-(\theta_3+2\beta-1)\zeta\eta-\frac{(\theta_2+1)t_2(t_1-1)\zeta}{t_1-t_2}+\frac{(\theta_1+1)t_1(t_2-1)\eta}{t_1-t_2}]V'_{\zeta}+$$
$$[\beta(\beta+\theta_3)\eta+(t_2-1)\theta_2\alpha+t_2\theta_2\beta]V.
\end{equation}

За счет операторных соотношений 
\begin{equation}\label{hein}\frac{\partial}{\partial\zeta}\zeta-\zeta\frac{\partial}{\partial\zeta}=1,\qquad \frac{\partial}{\partial\eta}\eta-\eta \frac{\partial}{\partial\eta}=1 \end{equation}
эти уравнения символически можно записать   как    ``квантования'' ($\varepsilon=1$)
\begin{equation}\label {qugomp}\varepsilon \frac{\partial V}{\partial t_i}=H_{Gar,t_i}(t_1,t_2,\zeta,\eta,-\varepsilon\frac{\partial}{\partial \zeta},-\varepsilon\frac{\partial}{\partial \eta})V \qquad{i=1,2},\end{equation}
определяемые гамильтонианами (\ref{Hkanasa}) полиномиальной системы Гарнье.

{\Large Замечание6.} За счет соотношений (\ref{hein}) уравнения (\ref{quoopp}),(\ref{quotpp}) можно записать и в виде ($\varepsilon=1$)
\begin{equation}\label {qugompp}\varepsilon \frac{\partial V}{\partial t_i}=H_{Gar,t_i}(t_1,t_2,\zeta,\eta,\varepsilon\frac{\partial}{\partial \zeta},\varepsilon\frac{\partial}{\partial \eta})V \qquad{i=1,2}.\end{equation}

Таким образом, замены (\ref{qdno}), (\ref{chenqz}) и конструкции разделов 2 и 3 дают решения ``квантований'' (\ref{qugomp})  полиномиальной системы Гарнье (\ref{Kanaso}). Эти решения
явным образом выписаны через решения совместных уравнений ИДМ (\ref{sch1}) с фиксацией (\ref{yard}). При этом коэффициенты  этих уравнений ИДМ однозначно выражаются (также явным образом) через множество совместных решений  систем (\ref{Kanaso}).

\bigskip

{\bf Приложение Случай $A_\infty=0$.}

\bigskip

{\bf 1.} Рассмотрим cлучай системы  уравнений (\ref{sch1})  
$$\Phi_x=A(x)\Phi=(\frac{A_1}{x-t_1}+\frac{A_2}{x-t_2}+\frac{A_3}{x-t_3}+\frac{A_4}{x-t_4})\Phi,
\quad\Phi_{t_i}=-\frac{A_i}{x-t_i}\Phi$$
с $A_1+A_2+A_3+A_4=A_\infty=0$. В силу последнего равенства 
$$A(x)=\frac{P_1x^2+P_2x+P_3}{(x-t_1)(x-t_2)(x-t_3)(x-t_4)}.$$

Все совместные решения этого случая системы (\ref{sch1}), удовлетворяя соотношениям
$$\sum\Phi_{t_i}=-\Phi_x,\quad\sum t_i\Phi_{t_i}=-x\Phi_x,\quad
x^2\Phi_x+\sum t_i^2\Phi_{t_i}=P_1\Phi,$$
имеют вид $\Phi=F(r,t,\xi)$, где
$$r=\frac{x-t_4}{x-t_3}:\frac{t_2-t_4}{t_2-t_3},\quad
t=\frac{t_1-t_4}{t_1-t_3}:\frac{t_2-t_4}{t_2-t_3},\quad
\xi=\frac{t_2-t_4}{t_2-t_3},$$
а функция $F$ есть совместное решение системы уравнений
$$F_r=\Big(\frac{a_1(\xi,t)}{r}+\frac{a_2(\xi,t)}{r-1}+\frac{a_3(\xi,t)}{r-t}\Big)F,\quad
F_t=\Big(-\frac{a_3(\xi,t)}{r-t}+\frac{a_3(\xi,t)}{\xi-t}\Big)F, F_\xi=Q(\xi,t)F.$$
После калибровочного преобразования $F=g(\xi,t)L$ с такой матрицей
$g(\xi,t)$, что $g_\xi=Qg$, получаем систему
$$L_r=\Big(\frac{b_1(t)}{r}+\frac{b_2(t)}{r-1}+\frac{b_3(t)}{r-t}\Big)L,\quad
L_t=-\frac{b_3(t)}{r-t}L,$$
которой удовлетворяет матрица $L=L(r,t)$, уже не зависящая от $\xi$.
Известно \cite[формулы (13), (14), (36) \S 2]{Fed}, \cite[главы 17 и 18]{Boli},\cite [раздел 3]{Mou}, что условие совместности этой системы сводится к шестому ОДУ Пенлеве. 

В случае $A_\infty=0$ также $\Delta_\infty=0$, и к уравнениям (\ref{odn})
добавляется еще одно, которое совместно с ними и с уравнениями (\ref{BPZ}):
$$\sum_{i=1}^mt_i^2Y'_{t_i}+x^2Y'_x+y^2Y'_y=-\Big(1+\sum_{i=1}^m\frac{\theta_i}{2}\Big)(\theta_1t_1+\theta_2t_2+\theta_3t_3+\theta_4t_4+x+y)Y$$
Это уравнение прямо следует из системы (\ref{sch1}). При $m=4$ 
совместное решение этих уравнений имеет вид $Y=\alpha(t,\xi)G(r,s,t)$, где 
$$s=\frac{y-t_4}{y-t_3}:\frac{t_2-t_4}{t_2-t_3}.$$  
Из формулы (\ref{NovDP}) несложно заключить, что $M=\tau(t,\xi)F(s,t,\xi)^{-1}F(r,t,\xi)$. Этот факт согласуется с предположением о том, что вышеупомянутое колибровочное преобразование $F=g(\xi,t)L$ задается матрицей  $g(\xi,t)=CL(\xi,t)^{-1}$: при $C=1$ имеем равенство  $M=\tau(t,\xi)L(s,t)^{-1}L(r,t)$. Но для уточнения вопроса о справедливости этого предположения нужно исходить из постановки задачи Римана - Гильберта и ее сведения к интегральным уравнениям. По всей видимости, формулы Фредгольма дают именно такой ответ.

Работа над статьей второго из авторов выполнена при поддержке  РНФ (грант 14-01-00171).

\end{document}